\documentclass[prc,twocolumn,twoside,showpacs,nofootinbib,floatfix]{revtex4}

\usepackage{graphicx,color,rotating,pifont}
\usepackage{amsmath,amssymb,bm}
\usepackage{ae}

\DeclareMathSymbol{\NS}{\mathord}{AMSb}{"4E}

\newcommand{\ket}[1]{\ensuremath{\,|{#1}\rangle}}
\newcommand{\braket}[2]{\ensuremath{\langle{#1}|{#2}\rangle}}
\newcommand{\matrixe}[3]{\ensuremath{\langle{#1}|\,{#2}\,|{#3}\rangle}}

\newcommand{\dmatrixe}[2]{\matrixe{#1}{#2}{#1}}

\newcommand{\trace}{\mathrm{tr}\,}

\newcommand{\expect}[1]{\ensuremath{\langle{#1}\rangle}}

\newcommand{\comm}[2]{\ensuremath{[{#1},{#2}]}}

\newcommand{\op}[1]{\ensuremath{#1}}
\newcommand{\adj}[1]{\ensuremath{{{#1}}^{\dag}}}
\newcommand{\corr}[1]{\ensuremath{\widetilde{#1}}}

\renewcommand{\vec}[1]{\ensuremath{\bm{#1}}}

\newcommand{\sixjsymb}[6]{\ensuremath{\left\{ \begin{matrix} 
  #1 & #2 & #3 \\ #4 & #5 & #6 \end{matrix} \right\}}}

\newcommand{\partd}[2]{\ensuremath{ \frac{\partial {#1}}{\partial {#2}} }}

\newcommand{\cO}{\ensuremath{\op{c}}}
\newcommand{\gO}{\ensuremath{\op{g}}}
\newcommand{\qO}{\ensuremath{\op{q}}}
\newcommand{\rO}{\ensuremath{\op{r}}}
\newcommand{\vO}{\ensuremath{\op{v}}}

\newcommand{\betaO}{\ensuremath{\op{\beta}}}
\newcommand{\etaO}{\ensuremath{\op{\eta}}}

\newcommand{\ccO}{\ensuremath{\adj{\op{c}}}}

\newcommand{\bbetaO}{\ensuremath{\adj{\op{\beta}}}}

\newcommand{\CO}{\ensuremath{\op{C}}}
\newcommand{\HO}{\ensuremath{\op{H}}}
\newcommand{\OO}{\ensuremath{\op{O}}}
\newcommand{\PO}{\ensuremath{\op{P}}}
\newcommand{\TO}{\ensuremath{\op{T}}}
\newcommand{\UO}{\ensuremath{\op{U}}}
\newcommand{\VO}{\ensuremath{\op{V}}}

\newcommand{\CCO}{\ensuremath{\adj{\op{C}}}}
\newcommand{\UUO}{\ensuremath{\adj{\op{U}}}}

\newcommand{\rV}{\ensuremath{\vec{r}}}

\newcommand{\nablaV}{\ensuremath{\vec{\nabla}}}
\newcommand{\lnablaV}{\ensuremath{\stackrel{\leftarrow}{\vec{\nabla}}}}
\newcommand{\rnablaV}{\ensuremath{\stackrel{\rightarrow}{\vec{\nabla}}}}

\newcommand{\EC}{\ensuremath{\mathcal{E}}}
\newcommand{\HC}{\ensuremath{\mathcal{H}}}

\newcommand{\pOV}{\ensuremath{\vec{\op{p}}}}
\newcommand{\qOV}{\ensuremath{\vec{\op{q}}}}
\newcommand{\rOV}{\ensuremath{\vec{\op{r}}}}

\newcommand{\LOV}{\ensuremath{\vec{\op{L}}}}

\newcommand{\ROV}{\ensuremath{\vec{\op{R}}}}

\newcommand{\sigmaOV}{\ensuremath{\vec{\op{\sigma}}}}

\newcommand{\Tint}{\ensuremath{\TO_\text{int}}}
\newcommand{\Tpair}{\ensuremath{\TO_\text{pair}}}
\newcommand{\Hint}{\ensuremath{\HO_\text{int}}}

\newcommand{\sigmasigmaO}{\ensuremath{\vec{\op{\sigma}}_{\!1}\!\cdot\!\vec{\op{\sigma}}_{\!2}}}

\newcommand{\tensorRQO}{\ensuremath{\op{S}_{12}(\rOV,\qOV_{\Omega})}}

\newcommand{\spinorbitO}{\ensuremath{(\vec{\op{L}}\cdot\vec{\op{S}})}}

\newcommand{\Rp}{\ensuremath{R_+}}

\newcommand{\Vlowk}{\ensuremath{V_{\text{low-k}}}}
\newcommand{\UCOM}{\ensuremath{\text{UCOM}}}
\newcommand{\VUCOM}{\ensuremath{V_{\text{UCOM}}}}

\newcommand{\aHO}{\ensuremath{a_{\text{HO}}}}
\newcommand{\eMax}{\ensuremath{e_{\text{max}}}}

\newcommand{\half}{\ensuremath{\tfrac{1}{2}}}

\newcommand{\nuc}[2]{\ensuremath{^{#2}\mathrm{#1}}}

\newcommand{\fm}{\ensuremath{\,\text{fm}}}

\newcommand{\keV}{\ensuremath{\,\text{keV}}}
\newcommand{\MeV}{\ensuremath{\,\text{MeV}}}

\newcommand{\linemediumsolid}[1][black]{\unitlength1ex 
  ({\color{#1}\begin{picture}(6,1)
  \linethickness{0.4mm}
  \put(0,0.5){\line(1,0){6.0}}
  \end{picture}})\nolinebreak
}

\newcommand{\symboldiamond}[1][black]{({\color{#1}$\blacklozenge$})}
\newcommand{\symboltriangle}[1][black]{({\color{#1}$\blacktriangle$})}
\newcommand{\symbolbox}[1][black]{({\color{#1}$\blacksquare$})}
\newcommand{\symbolcircle}[1][black]{({\color{#1}\ding{108}})}

\newcommand{\symbolboxopen}[1][black]{({\color{#1}$\square$})}
\newcommand{\symbolcircleopen}[1][black]{({\color{#1}\ding{109}})}
\newcommand{\symbolxcross}[1][black]{({\color{#1}\ding{54}})}

\definecolor{FGViolet}{rgb}{0.61,0.32,0.61}
\definecolor{FGDarkBlue}{rgb}{0,0,0.6}
\definecolor{FGBlue}{rgb}{0,0,0.8}
\definecolor{FGLightBlue}{rgb}{0.2, 0.6, 0.8}
\definecolor{FGGreen}{rgb}{0.2,0.7,0.2}
\definecolor{FGLightGreen}{rgb}{0.4,1,0.4}
\definecolor{FGYellow}{rgb}{1,0.95,0}
\definecolor{FGOrange}{rgb}{0.95,0.5,0.1}
\definecolor{FGRed}{rgb}{0.8,0,0}
\definecolor{FGWhite}{rgb}{1,1,1}
\definecolor{FGLightGray}{rgb}{0.8,0.8,0.8}
\definecolor{FGGray}{rgb}{0.5,0.5,0.5}
\definecolor{FGDarkGray}{rgb}{0.3,0.3,0.3}
\definecolor{FGBlack}{rgb}{0,0,0}

\begin{document}

\title{Pairing in the framework of the unitary correlation operator method (UCOM): Hartree-Fock-Bogoliubov calculations}

\author{H. Hergert}
\email{Heiko.Hergert@physik.tu-darmstadt.de}

\author{R. Roth}
\email{Robert.Roth@physik.tu-darmstadt.de}

\affiliation{Institut f\"ur Kernphysik, Technische Universit\"at Darmstadt,
64289 Darmstadt, Germany}

\date{\today}

\begin{abstract}
In this first in a series of articles, we apply effective interactions derived by the Unitary Correlation Operator Method (UCOM) to the description of open-shell nuclei, using a self-consistent Hartree-Fock-Bogoliubov framework to account for pairing correlations. To disentangle the particle-hole and particle-particle channels and assess the pairing properties of $\VUCOM$, we consider hybrid calculations using the phenomenological Gogny D1S interaction to derive the particle-hole mean field. In the main part of this article, we perform calculations of the tin isotopic chain using $\VUCOM$ in both the particle-hole and particle-particle channels. We study the interplay of both channels, and discuss the impact of non-central and non-local terms in realistic interactions as well as the frequently used restriction of pairing interactions to the ${}^1S_0$ partial wave. The treatment of the center-of-mass motion and its effect on theoretical pairing gaps is assessed independently of the used interactions.
\end{abstract}


\maketitle

\clearpage

\section{Introduction}
Recent years have seen a revival of nuclear structure physics, motivated by new experimental advances in the use of radioactive beams in existing and proposed facilities at GSI/FAIR, RIKEN, GANIL, and other laboratories worldwide, as well as new theoretical approaches to the nuclear many-body problem. The application of effective field theory (EFT) and renormalization group (RG) methods has provided systematic approaches to the construction of effective nuclear interactions that maintain a stringent link to QCD, either directly on the formal level or by the reproduction of low-energy observables like NN phase shifts and deuteron properties. The former have culminated in the derivation of a consistent set of two- and higher many-nucleon interactions in the framework of chiral EFT at next-to-next-to-next-to-leading order (N3LO) \cite{Epelbaum:2005pn,Entem:2003ft}, while the latter have revealed the universal aspects of realistic NN interactions by decoupling low- and high-momentum modes via RG decimations in the case of $\Vlowk$ \cite{Bogner:2003wn}, or unitary transformations in the Similarity Renormalization Group (SRG) \cite{Bogner:2006pc}.

While starting from a different premise, i.e., the explicit treatment of correlations induced by the repulsive core and the tensor force of realistic NN interactions like Argonne V18 or CD-Bonn (see \cite{Machleidt:2001rw} for a review), the Unitary Correlation Operator Method (UCOM) \cite{Feldmeier:1997zh,Neff:2002nu} shares many characteristics of the RG-derived low-momentum interactions. This is particularly true for the SRG approach, where the dynamical generator of the unitary transformation is related to the generators of the UCOM transformation \cite{Hergert:2007wp,Roth:2008km}.

In our previous works, correlated interactions derived in the UCOM framework, referred to as $\VUCOM$ in the following, have proven their merit in a wide range of many-body methods, from ab-initio calculations of light nuclei in the No Core Shell Model \cite{Roth:2005pd, Roth:2007sv, Roth:2009cw} to Hartree-Fock (HF) and HF-based extensions like Many-Body Perturbation Theory \cite{Roth:2005ah,Guenther:2009}, RPA \cite{Paar:2006ua, Papakonstantinou:2006vc}, and Second RPA \cite{Papakonstantinou:2008sf}. Since HF-based approaches do not account for pairing correlations, they are expected to work best for closed-shell nuclei. The aim of this first in a series of papers is the extension of our calculations to open-shell nuclei by constructing a self-consistent Hartree-Fock-Bogoliubov (HFB) framework (see, e.g., \cite{Ring:1980}).

Initial attempts to solve the HFB equations using (at the time) ``realistic'' interactions \cite{Bartouv:1969zz, Goodman:1970mh} were hampered by the strong repulsive core of the NN interaction, which leads to infinities in mean-field methods because Slater determinants are inherently unable to describe the required correlations. Brueckner's $G$-matrix approach \cite{Day:1967zz} provided a way to deal with this problem by resumming particle-particle ladder diagrams, leading to a well-behaved ``tamed'' interaction, but remained problematic, e.g., due to the starting-energy dependence. At the same time, it was observed that a similar resummation was not required to regularize the NN gap equation (see, e.g., \cite{Henley:1964zz,Kennedy:1964zz}), and, in modern terms, the ``bare'' interaction could be used directly, prompting a disparate treatment of the particle-hole and particle-particle channels in self-consistent field calculations in the following decades.

Parallel to these original Hamiltonian-based approaches, Negele and Vautherin introduced the Density Matrix Expansion \cite{Negele:1972zp} in nuclear physics. While their work was tied to the Hamiltonian-based approaches by using similar concepts as in $G$-matrix methods, it also provided a foundation for the form of Skyrme-type energy functionals, and paved the way for phenomenological Density Functional Theory (DFT), which became the standard framework for self-consistent field methods until today (see, e.g., \cite{Bender:2003jk} for a comprehensive review). While current phenomenological density functionals are able to describe nuclear bulk properties like binding energies and charge radii with high accuracy near the valley of stability, they often perform inadequately in the description of spectroscopic observables or exotic nuclei. As a result, considerable theoretical effort is under way to improve the phenomenological functionals (see, e.g., \cite{Zalewski:2008is,Duguet:2008rr}), or to \emph{construct} it from first principles by applying EFT methods \cite{Furnstahl:2007xm}. Given the guiding principles of EFTs with respect to consistency, one then has to ask whether one should actually demand the treatment of the particle-hole and particle-particle channels on the same footing --- especially since effective interactions derived in the (S)RG or the UCOM approaches no longer require resummations in the particle-hole channel.

A strong argument in favor of such consistency was encountered in phenomenological DFT in recent years. Many-body methods beyond the mean field, e.g., Generator Coordinate approaches, involve configuration mixing of non-orthogonal Slater determinants, and rely on subtle cancellations between singular terms in the particle-hole and particle-particle channel (see, e.g., \cite{Anguiano:2001in,Dobaczewski:2007ch}). A Hamiltonian provides an ideal starting point for these methods, because the use of the same interaction in both channels automatically guarantees these cancellations, whereas one has to go to some lengths to implement a ``regularization'' scheme in DFT to remove spurious contributions to the energy and other expectation values due to the use of separate particle-hole and pairing functionals \cite{Lacroix:2008rj,Bender:2008rn}. 

For the reasons discussed above, the aim of this paper is the formulation of a fully self-consistent HFB scheme based on an intrinsic Hamiltonian, using a family of correlated interactions based on the realistic Argonne V18 interaction \cite{Wiringa:1994wb} in both interaction channels. After briefly reviewing the basics of the HFB approach and the Unitary Correlation Operator Method in Sect. \ref{sec:form}, we discuss certain details of our implementation in Sect. \ref{sec:implement}. This includes a comparison of the convergence behavior of $\VUCOM$ with the phenomenological Gogny D1S interaction \cite{Berger:1991}, as well as a detailed discussion of the center-of-mass treatment. In Sect. \ref{sec:pp_ucom}, we investigate the properties of $\VUCOM$ as a pairing interaction, using the Gogny force to generate the mean field and thereby disentangle the particle-hole and particle-particle channels. This section also includes a comparison with the pairing properties of SRG-evolved interactions. Section \ref{sec:ucom} presents results from fully self-consistent HFB calculations with $\VUCOM$. 

\section{\label{sec:form}Formalism}
\subsection{\label{sec:hfb}Hartree-Fock-Bogoliubov Theory}
The HFB approach \cite{Ring:1980} aims for a simultaneous mean-field description of the particle-hole and particle-particle channels of the NN interaction by introducing quasiparticle operators $\{\betaO_k, \bbetaO_k\}_{k\in\NS}$ via the Bogoliubov transformation 
\begin{align}\label{eq:bt}
  \bbetaO_k &= \sum_l U_{lk}\ccO_l + V_{lk}\cO_l\,,\\
  \betaO_k  &= \sum_l U_{lk}^*\cO_l +V_{lk}^*\ccO_l\,,
\end{align}
where $\cO_k$ and $\ccO_k$ are annihilation and creation operators in the standard particle space. The Bogoliubov conditions  
\begin{subequations}\label{eq:bt_cond}
 \begin{align}
   \adj{U}U+\adj{V}V &= 1\,,  & U\adj{U} + V^*V^T&=1\,, \\
   U^TV + V^TU &=0\,,         & U\adj{V} + V^*U^T&=0\,,
 \end{align}
\end{subequations}
ensure that the $\{\betaO_k, \bbetaO_k\}_{k\in\NS}$ satisfy the canonical anticommutation relations. In HFB approximation, the nuclear ground state $\ket{\Psi}$ is defined (up to a unitary transformation) by the \emph{quasiparticle vacuum},
\begin{equation}\label{eq:qp_vacuum}
  \betaO_k\ket{\Psi} = 0\,.
\end{equation}

Taking account of the center-of-mass contribution to the kinetic energy, we introduce the intrinsic kinetic energy operator
\begin{equation}\label{eq:def_Tint}
  \Tint= \frac{2}{A}\sum_{i<j}^A \frac{\qOV^2_{ij}}{2\mu}\,,\quad \mu=\frac{m_N}{2}\,,
\end{equation}
where
\begin{equation}
  \qOV_{ij}=\frac{1}{2}(\pOV_i-\pOV_j)
\end{equation}
is the relative momentum and $\mu$ the reduced mass, given in terms of the nucleon mass $m_N$. Using $\Tint$, the intrinsic many-body Hamiltonian reads
\begin{equation}
  \Hint=\Tint+\VO = \TO-\TO_\text{cm}+\VO\,.
\end{equation}
The intrinsic energy of the HFB ground state can then be expressed in terms of the density matrix 
\begin{equation}\label{eq:def_rho}
  \rho_{kk'} = \dmatrixe{\Psi}{\ccO_{k'}\cO_k}=\left(V^*V^T\right)_{kk'}
\end{equation}
and the pairing tensor
\begin{equation}\label{eq:def_kappa}
  \kappa_{kk'} = \dmatrixe{\Psi}{\cO_{k'}\cO_k}=\left(V^*U^T\right)_{kk'}\,
\end{equation}
as
\begin{equation}\label{eq:def_energy}
   E[\rho,\kappa,\kappa^*] = \frac{1}{2}\trace(h\rho) - \frac{1}{2}\trace(\Delta\kappa^*)\,,
\end{equation}
where the hermitian particle-hole field
\begin{equation}
  h_{kk'}\equiv\partd{E}{\rho_{k'k}}\equiv \sum_{qq'}\left(\frac{2}{A}\bar{t}+\bar{v}\right)_{kq'k'q}\rho_{qq'}\label{eq:def_gamma}
\end{equation}
and the antisymmetric pairing field
\begin{equation}
  \Delta_{kk'}\equiv\partd{E}{\kappa_{kk'}^*}=\frac{1}{2}\sum_{qq'}\left(\frac{2}{A}\bar{t}+\bar{v}\right)_{kk'qq'}\kappa_{qq'}
  \label{eq:def_delta}
\end{equation}
have been introduced. $\bar{t}$ and $\bar{v}$ denote the antisymmetrized matrix elements of the intrinsic kinetic energy and the NN interaction, respectively. 

The HFB ground state is obtained by performing a variation of the energy with respect to $\rho$ and $\kappa$, subject to the constraint
\begin{equation}\label{eq:number_constraint}
   \trace \rho = N\,,
\end{equation}
which ensures the conservation of the mean particle number. Carrying out the variation, one obtains the Hartree-Fock-Bogoliubov equations
\begin{equation}\label{eq:hfb_equations}
  \HC\begin{pmatrix}
    U_k \\ V_k
  \end{pmatrix}
  \equiv
  \begin{pmatrix}
    h-\lambda & \Delta       \\
   -\Delta^* & -h^* + \lambda 
  \end{pmatrix}
  \begin{pmatrix}
    U_k \\ V_k
  \end{pmatrix}
  = E_k
  \begin{pmatrix}
    U_k \\ V_k
  \end{pmatrix}\,,
\end{equation}
where we have introduced the HFB Hamiltonian $\HC$. Due to the use of an intrinsic Hamiltonian, the Lagrange multiplier $\lambda$ in Eq. \eqref{eq:hfb_equations} can no longer be identified directly with the Fermi energy of the system (see Sect. \ref{sec:kinetic_int}). Equation \eqref{eq:hfb_equations} constitutes an eigenvalue problem that has to be solved self-consistently due to the dependence of $\HC$ on $\rho$ and $\kappa$.

\subsection{Spherical Symmetry}
Assuming spherical symmetry, the Bogoliubov transformation reduces to the form
\begin{align}\label{eq:bt_sho}
  \bbetaO_{nljm} &= \sum_{n'}U^{(lj)}_{n'n}\ccO_{n'ljm} + (-1)^{j+m}V^{(lj)}_{n'n}\cO_{n'lj-m}\,,\\
  \betaO_{nljm}  &= \sum_{n'}U^{(lj)}_{n'n}\cO_{n'ljm} + (-1)^{j+m}V^{(lj)}_{n'n}\ccO_{n'lj-m}\,,
\end{align}
where $n$ is a radial quantum number and the upper indices $(lj)$ mark the (diagonal) angular-momentum quantum numbers. Aside from the explicit phase factor in Eq. \eqref{eq:bt_sho}, the transformation is independent of the angular-momentum projection $m$. Using the matrices $U$ and $V$ from Eq. \eqref{eq:bt_sho}, one can define the reduced matrices $\rho^{(lj}_{nn'}$ and $\kappa^{(lj)}_{nn'}$, 
\begin{align}
	\rho_{nljm,n'l'j'm'} &= \left[VV^T\right]^{(lj)}_{nn'}\delta_{jj'}\delta_{ll'}\delta_{mm'}\notag\\
			    &\equiv \rho^{(lj)}_{nn'}\delta_{jj'}\delta_{ll'}\delta_{mm'}\,, \label{eq:def_rholj}\\
	\kappa_{nljm,n'l'j'm'} &= (-1)^{j-m}\left[VU^T\right]^{(lj)}_{nn'}\delta_{jj'}\delta_{ll'}\delta_{m-m'}\notag\\
			    &\equiv (-1)^{j-m}\kappa^{(lj)}_{nn'}\delta_{jj'}\delta_{ll'}\delta_{m-m'}\,,
	\label{eq:def_kaplj}
\end{align}
which are both symmetric and real. The antisymmetry of the pairing tensor is now contained entirely in the phase factor. Correspondingly, the reduced fields are defined by
\begin{equation}
  \Gamma_{nljm,n'l'j'm'} 
	\equiv \delta_{jj'}\delta_{ll'}\delta_{mm'}\Gamma^{(lj)}_{nn'}\label{eq:def_gamlj}
\end{equation}
and
\begin{equation}
     \Delta_{nljm,n'l'j'm'}
     \equiv\delta_{jj'}\delta_{ll'}\delta_{m,-m'}(-1)^{j-m}\Delta^{(lj)}_{nn'}\,.
     \label{eq:def_dellj}
\end{equation}

With these definitions, the reduced HFB equations read
\begin{equation}\label{eq:hfb_lj}
  \begin{pmatrix}
    h^{(lj)}-\lambda & -\Delta^{(lj)}       \\
   -\Delta^{(lj)} & -h^{(lj)} + \lambda 
  \end{pmatrix}
  \begin{pmatrix}
    U_k^{(lj)} \\ V_k^{(lj)}
  \end{pmatrix}
  = E_k
  \begin{pmatrix}
    U_k^{(lj)} \\ V_k^{(lj)}
  \end{pmatrix}\,.
\end{equation}
\subsection{\label{sec:canon}Canonical Basis}
The canonical basis is a convenient tool for the discussion of the HFB problem, because the HFB equations \eqref{eq:hfb_equations} assume the same form as in the Bardeen-Cooper-Schrieffer (BCS) case in this representation. It is obtained by diagonalizing the one-body density matrix \eqref{eq:def_rho}, whose eigenvalues $v^2_\mu$ are interpreted as occupation probabilities of the corresponding canonical states $\ket{\mu}$. The accompanying coefficients $u^2_\mu$ are defined up to a phase by the condition [cf. Eq. \eqref{eq:bt_cond}]
\begin{equation}
  u_\mu^2 + v_\mu^2 = 1\,.
\end{equation}
Analogous to the BCS case one can then define generalized single-particle energies and state-dependent gaps (see, e.g., \cite{Dobaczewski:1995bf}) via the matrix elements
\begin{align}
  \epsilon_\mu &= h_{\mu\mu}\,,\\
  \Delta_{\mu} &= \Delta_{\mu\bar\mu}\,,
\end{align}
as well as the canonical quasiparticle energy 
\begin{equation}\label{eq:def_eqpcan}
  \EC_{\mu} = \sqrt{(\epsilon_\mu - \lambda)^2 + \Delta_\mu^2}\,.
\end{equation}
Here, $\ket{\bar\mu}$ is the canonical conjugate state of $\ket{\mu}$ (e.g., the time-reversed state in systems with time-reversal symmetry). $\EC_\mu$ is just the \emph{diagonal matrix element} of the HFB Hamiltonian in the canonical basis, and therefore generally \emph{not identical} to any of the quasiparticle energies obtained by diagonalizing $\HC$. In terms of the newly defined quantities, the canonical coefficients can be written as
\begin{gather}\label{eq:def_ucan}
  u_\mu = \sqrt{\frac{1}{2}\left(1+\frac{\epsilon_\mu-\lambda}{\EC_\mu}\right)}\,,\\
  v_\mu = \mathrm{sgn}\,(\Delta_\mu)\sqrt{\frac{1}{2}\left(1-\frac{\epsilon_\mu-\lambda}{\EC_\mu}\right)}\label{eq:def_vcan}\,.
\end{gather}
where we have adopted the phase conventions of \cite{Dobaczewski:1995bf}.

\subsection{Gap Definitions}
Experimentally, the odd-even staggering of nuclear binding energies provides a clear signal of pairing correlations in the finite nucleus. This staggering is analyzed via the ground-state energy differences of several neighboring nuclei. Recently, it was argued that the odd-centered three-point formula
\begin{equation}\label{eq:def_expgap}
  \Delta^{(3)}(N)=-\frac{1}{2}\left(E(N+1)-2 E(N) + E(N-1)\right)\,,
\end{equation}
provides the clearest measure of pairing correlations along isotopic (or isotonic) chains, because it is least affected by particle-hole effects (see, e.g., \cite{Duguet:2001gs}, which also discusses further refinements). The best way to compare these experimental ``gaps'' to theory would be the application of Eq. \eqref{eq:def_expgap} to theoretical ground-state energies. Since the treatment of odd nuclei in a HFB framework requires further approximations with respect to the blocking of levels by the unpaired nucleon, we defer such calculations to the future. We point out, however, that such calculations are facilitated in a Hamiltonian-based approach because the interaction is already completely determined.

While the pairing energy in Eq. \eqref{eq:def_energy} provides an obvious measure of pairing correlations in a theoretical calculation, it cannot be related directly to the experimental gap \eqref{eq:def_expgap}. Thus, one usually turns to the canonical basis, where theoretical gaps that allow some form of comparison can be defined. In analogy to BCS theory, one can then consider the state-dependent gap of the canonical state with the \emph{lowest quasiparticle energy} (cf. Sect. \ref{sec:canon}) as a measure of pairing correlations (see, e.g., \cite{Duguet:2007be,Lesinski:2008cd}):
\begin{equation}\label{eq:def_canDelta}
   \Delta = \Delta_{\mu_0},\quad \EC_{\mu_0}=\min_\mu \EC_\mu\,.
\end{equation}
Various other prescriptions for the gap are used in the literature as well, in particular 
\begin{equation}\label{eq:def_avgDelta}
   \expect{\Delta}=\frac{\sum_\mu u_\mu v_\mu\Delta_\mu}{\sum_\mu u_\mu v_\mu}\,,
\end{equation}
which corresponds to the average of the pairing energy over the paired canonical states at the Fermi surface \cite{Bender:2003jk}. 
To interpret our theoretical results, we will primarily use the canonical gap \eqref{eq:def_canDelta}. In comparison, the averaged gap has only slightly different values and exhibits somewhat smoother trends. Any exceptions to this behavior will be addressed explicitly in the discussion.

\subsection{Unitary Correlation Operator Method (UCOM)}
The Unitary Correlation Operator Method is motivated by physical considerations on the structure and origin of the dominant many-body correlations. The short-range repulsion in the central part of the NN interaction drives the interacting nucleon pair apart. The tensor interaction induces correlations between the relative distance and the spin of the nucleon pair, leading to the characteristic mixing between components with relative orbital angular momentum $L$ and $L\pm2$ in the $S=1$ channel. To imprint these correlations on a many-body state, we construct a unitary transformation with the generators
\begin{equation}\label{eq:gen_r}
  \gO_r = \frac{1}{2}\left(\qO_r s(r) + s(r)\qO_r\right)
\end{equation}
and
\begin{align}\label{eq:gen_tens}
  \gO_\Omega &= \vartheta(r)\tensorRQO \notag \\
   &\equiv\vartheta(r)\frac{3}{2}
         \left(\left(\sigmaOV_1\!\cdot\!\qOV_{\Omega}\right)\left(\sigmaOV_2\!\cdot\!\rOV\right)+\right.         
         \left.\left(\sigmaOV_1\!\cdot\!\rOV\right)\left(\sigmaOV_2\!\cdot\!\qOV_{\Omega}\right)\right)\,,
\end{align}
where 
\begin{gather}
  \qO_r \equiv \frac{1}{2}\left(\qOV\cdot\frac{\rOV}{r} + \frac{\rOV}{r}\cdot\qOV\right)\,,\\
  \qOV_\Omega \equiv \qOV-\qO_r\frac{\rOV}{r}=\frac{1}{2r^2}\left(\LOV\times\rOV-\rOV\times\LOV\right)\,.
\end{gather}
The generator $\gO_r$ uses the radial part of the relative momentum operator $\qOV$ to create a shift in the radial direction, while $\gO_\Omega$ is constructed from the so-called orbital momentum, i.e., the angular part of $\qOV$, and generates shifts perpendicular to $\rOV$. The strength and range of the transformation is governed by the shift function $s(r)$ and the tensor correlation function $\vartheta(r)$. Rather than using the shift function directly, it is more practical to define the central correlation function $\Rp(r)$ via the integral equation
\begin{equation}
  \int_{r}^{\Rp(r)}\frac{d\xi}{s(\xi)}=1\,,
\end{equation}
which implies $\Rp(r)\approx r+s(r)$ for a weakly $r$-dependent $s(r)$.

The unitary transformation is then written as 
\begin{equation}\label{eq:simtrans}
  \CO\equiv\CO_\Omega\CO_r\equiv\exp\bigg(\!-i\sum_{j<k}\gO_{\Omega,jk}\bigg)\!\exp\bigg(\!-i\sum_{j<k}\gO_{r,jk}\bigg)\,,
\end{equation}
where the sum runs over all nucleon pairs. One can now proceed to calculate expectation values either by applying $\CO$ to the many-body state $\ket{\Psi}$ or to a given observable $\OO$, yielding either a correlated state $\ket{\corr{\Psi}}$ or a correlated operator $\corr{\OO}$:
\begin{equation}
  \matrixe{\corr{\Psi}}{\OO}{\corr{\Phi}}=\matrixe{\Psi}{\CCO_r\CCO_\Omega\OO\CO_\Omega\CO_r}{\Phi}=
  \matrixe{\Psi}{\corr{\OO}}{\Phi}\,.
\end{equation}
The structure of the transformation \eqref{eq:simtrans} implies that $\corr{\OO}$ is an $A$-body operator in Fock space, which can be expressed in terms of irreducible contributions $\corr{\OO}^{[n]}$ for a specific particle number $n\leq A$ via the cluster expansion
\begin{equation}
  \corr{\OO}=\CCO\OO\CO=\corr{\OO}^{[1]}+\corr{\OO}^{[2]}+\ldots+\corr{\OO}^{[A]}\,.
\end{equation}
If the range of the correlation functions is small compared to the mean inter-particle distance, we can employ the two-body approximation and omit negligible cluster terms beyond the second order (for details see Refs. \cite{Feldmeier:1997zh, Neff:2002nu, Roth:2005pd}). For the construction of the correlated Hamiltonian in two-body approximation, it is then sufficient to consider the Hamiltonian in the two-nucleon system, 
\begin{equation}\label{eq:hrel}
  \Hint = \Tint+\VO \equiv \frac{\qOV^2}{2\mu} + \VO\,,
\end{equation}
where we have already subtracted the center-of-mass kinetic energy, which is not affected by the correlation procedure. Applying the correlation operators, 
\begin{equation}
  \CCO_r\CCO_\Omega\Hint\CO_\Omega\CO_r = \Tint + \corr{\TO}_\text{int}^{[2]}+\corr{\VO}^{[2]}+\ldots\,,
\end{equation}
and collecting the two-body contributions from the correlated kinetic energy and the transformed interaction, we obtain the effective interaction $\VO_{\UCOM}$: 
\begin{equation}
  \VO_\UCOM\equiv\corr{\TO}_\text{int}^{[2]}+\corr{\VO}^{[2]}\,.
\end{equation}
The evaluation of the matrix elements of $\VO_\UCOM$ in a partial-wave basis is discussed in detail in Ref. \cite{Roth:2005pd}.

\subsection{\label{sec:srg_ucom}SRG-generated $\VUCOM$}
In a recent pair of papers \cite{Hergert:2007wp, Roth:2008km}, we have studied the connection of the UCOM to the Similarity Renormalization Group (SRG) approach to the construction of effective NN interactions \cite{Bogner:2006pc}. There, the many-body Hamiltonian $\HO$ is evolved towards a block-diagonal structure in momentum space via the flow equation
\begin{equation} \label{eq:h_flow}
  \frac{d\HO_{\bar\alpha}}{d\bar\alpha}=\comm{\etaO(\bar\alpha)}{\HO_{\bar\alpha}}\,,\quad\HO_0=\HO\,,
\end{equation}
where $\bar\alpha$ denotes the flow parameter and
\begin{equation}
  \HO_{\bar\alpha}\equiv \UO(\bar\alpha)\HO\UUO(\bar\alpha)\equiv\Tint+\VO_{\bar\alpha}\,.
\end{equation}
The dynamical generator of the flow is defined by  
\begin{equation}\label{eq:gen}
  \etaO(\bar\alpha)=\comm{\Tint}{\HO_{\bar\alpha}}=\comm{\frac{\qOV^2}{2\mu}}{\HO_{\bar\alpha}}\,,
\end{equation}
where $\qOV$ is the relative momentum operator. Equation \eqref{eq:gen} is only appropriate for an evolution in two-body space, an assumption corresponding to the two-body approximation used in the UCOM framework. A generalization to three-nucleon or many-nucleon forces is straightforward in principle, but too demanding to allow the numerical evolution of realistic 3N Hamiltonians at present.

In Ref. \cite{Roth:2008km}, we describe a procedure by which a mapping between an uncorrelated trial state and the deuteron wave function of the SRG-evolved Hamiltonian $\HO_{\bar\alpha}$ defines a set of central and tensor correlation functions for use in the generators \eqref{eq:gen_r} and \eqref{eq:gen_tens}. Similar to the SRG-evolved $V_{\bar\alpha}$, the resulting $\VUCOM$ is uniquely determined by the parameter $\bar\alpha$ and the parent interaction, and offers a significantly enhanced convergence. At the same time, however, the saturation properties are quite different, because $\VUCOM$ does not produce the same strong overbinding for large nuclear masses as $V_{\bar\alpha}$ at the two-body level (see also Ref. \cite{Bogner:2007rx}).

For practical applications, we optimize the value of $\bar\alpha$ by considering No-Core Shell Model Calculations of the Tjon-line in $\nuc{H}{3}$ and $\nuc{He}{4}$. For $\bar\alpha=0.04\fm^4$, the resulting ground-state energies of these nuclei are close to the experimental values \emph{without} the inclusion of 3N forces --- in other words 3N forces generated by the UCOM transformation more or less cancel genuine 3N forces that would need to supplement the parent NN interaction in order to obtain the correct ground-state energies. 

\section{\label{sec:implement}Implementation}
\subsection{Basics}
\begin{figure}[t]
  \includegraphics[width=\columnwidth]{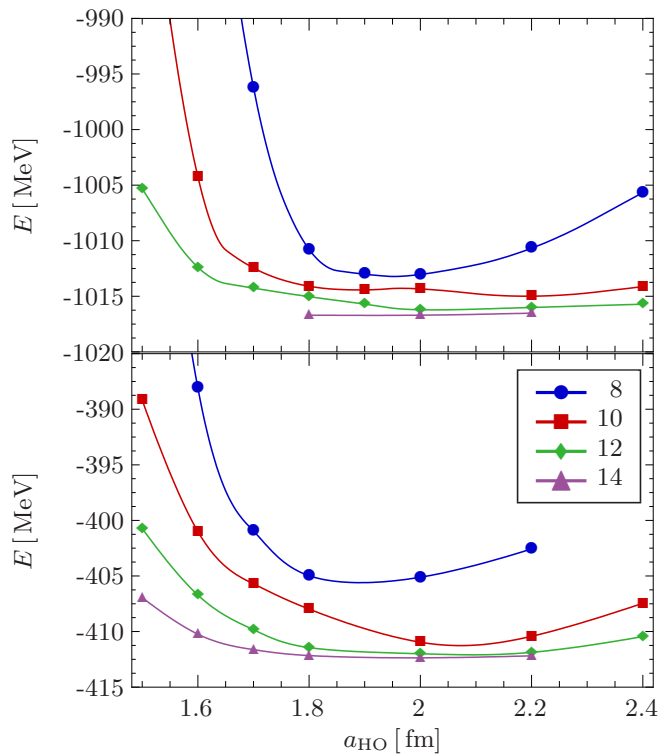}
  \caption{\label{fig:convergence}(Color online) Ground-state energy of $\nuc{Sn}{120}$ for various basis sizes $\eMax$ (see inset) and oscillator lengths $\aHO$, using the Gogny D1S interaction (top) and the SRG-optimized $\VUCOM$ with $\bar\alpha=0.4$ (bottom).}
\end{figure}

Our implementation of the HFB method makes use of the framework established for Hartree-Fock calculations with $\VUCOM$ \cite{Roth:2005pd, Roth:2005ah}. The eigensystem \eqref{eq:hfb_equations} or \eqref{eq:hfb_lj} is solved in a spherical harmonic oscillator (SHO) configuration space, using a truncation in the oscillator quantum number
\begin{equation}
  e = 2n + l\,,
\end{equation}
where $n$ and $l$ denote the radial oscillator quantum number and orbital angular momentum, respectively. An original implementation of the modified Broyden's method discussed in Ref. \cite{Johnson:1988} is employed to accelerate the convergence of the HFB fields (see \cite{Baran:2008bg} for recent applications in nuclear structure calculations).

Figure \ref{fig:convergence} illustrates the convergence of our HFB calculations for the sample nucleus $\nuc{Sn}{120}$, using an SRG-generated $\VUCOM$ with $\bar\alpha=0.4$ as well as the phenomenological Gogny D1S interaction \cite{Berger:1991}. The convergence rate is rather similar for both interactions as the single-particle basis size is increased. Since the typical energy gain by increasing $\eMax$ from $12$ to $14$ (corresponding to 13 or 15 major oscillator shells, respectively) is merely $1-2\,\MeV$ for the Gogny D1S interaction and even smaller for $\VUCOM$, we adopt the basis with $\eMax=12$ for the remainder of this work. Using this truncation, the residual dependence of the ground-state energy on the oscillator parameter $\aHO$ is already quite weak over a wide range of values --- nevertheless, we usually carry out calculations for a set of $\aHO$'s to explicitly minimize the energy in this respect as well.

Finally, we note that the converged ground-state energy obtained with $\VUCOM$ is in line with previous Hartree-Fock results \cite{Roth:2005ah, Roth:2008km}, providing less than half of the experimental binding energy. The missing binding energy is due to missing long-range correlations, which can be recovered by going beyond the mean-field approximation \cite{Roth:2005ah}, as well as omitted 3N or higher many-nucleon forces. In contrast, the Gogny D1S interaction is fit to experimental ground-state energies, providing the bulk of the $\nuc{Sn}{120}$ binding energy already in a mean-field calculation.

\subsection{\label{sec:kinetic_int}Intrinsic Kinetic Energy}
\begin{figure}[t]
  \includegraphics[width=\columnwidth]{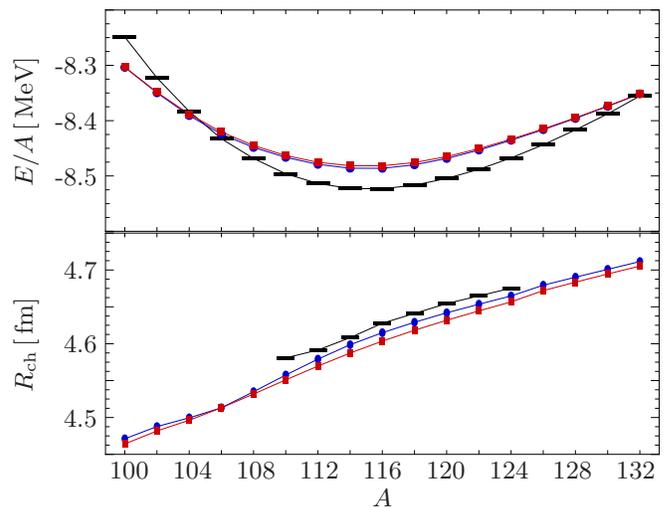}
  \caption{\label{fig:goge}(Color online) Ground-state energies (top) and charge radii (bottom) for the tin isotopes, calculated with the Gogny D1S interaction: full intrinsic kinetic energy \symbolcircle[FGBlue]\, and one-body approximation \symbolbox[FGRed]. Experimental values are indicated by black bars \cite{Audi:2002rp,DeVries:1987}.}
\end{figure}

An interesting issue that is rarely considered in the literature is the effect of the center-of-mass correction (see, however, the detailed study in \cite{Anguiano:2000dp}). As indicated in Sect. \ref{sec:hfb}, we formulate the HFB equations using the intrinsic kinetic energy, which 
can be expressed either in terms of two-body operators or a combination of one- and two-body terms: 
\begin{align}\label{eq:tint_tb}
  \Tint&=\frac{1}{2A}\sum_{i<j}\frac{(\pOV_i-\pOV_j)^2}{m}\\
                &=\left(1-\frac{1}{A}\right)\sum_i\frac{\pOV_i^2}{2m}-\frac{1}{Am}\sum_{i<j}\pOV_i\cdot\pOV_j\,.\label{eq:tint_obtb}
\end{align}
The use of an intrinsic Hamiltonian has a number of consequences for our calculations. While the total ground-state energy is lowered by the center-of-mass correction, the pairing field and pairing energy obtain positive contributions from the two-body part of the $\Tint$, leading to a reduction compared to calculations without center-of-mass corrections. In the plain Hartree-Fock case, the eigenvalues of the intrinsic HF Hamiltonian can be no longer directly identified with single-particle energies because its $A$-dependence invalidates Koopmans' theorem. Indeed, if the eigenvalues are interpreted as perturbative approximations to the exact separation energies,
\begin{equation}
  \varepsilon^{HF}_\mu\approx E_{N+1}-E_N\,,
\end{equation}
one sees that the $A$-dependence of the Hamiltonian leads to correction terms that account for rearrangement effects caused by the addition or removal of a nucleon (see Refs. \cite{Jaqua:1992, Roth:2005ah}). 

\begin{figure}[t]
  \includegraphics[width=\columnwidth]{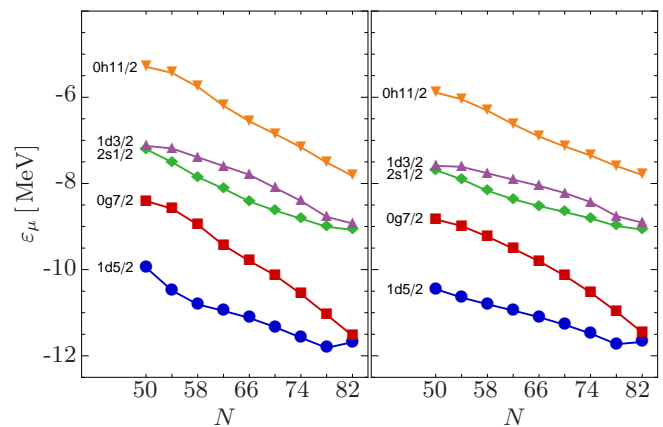}
  \caption{\label{fig:ecmcomp}(Color online) Canonical single-particle spectra of tin isotopes: full intrinsic kinetic energy (left) and one-body approximation (right). Calculations were done with the Gogny D1S interaction.}
\end{figure}

In the HFB case, the need for a similar procedure becomes evident in the behavior of the Lagrange parameter $\lambda$, which generally assumes \emph{positive} values and can therefore not directly be identified with the Fermi energy of the system. Unfortunately, the addition or removal of a particle to a system with pairing is a non-trivial issue \cite{Duguet:2001gr, Duguet:2001gs}, and the generalization of the aforementioned single-particle energy correction terms to the HFB case is not obvious, especially since part of the correction is state-dependent, whereas $\lambda$ is no longer associated with a definite single-particle level. At present, we therefore adopt the simpler state-independent correction discussed in Ref. \cite{Jaqua:1992},   
\begin{equation}\label{eq:def_lambda_corr}
  \lambda^\text{corr} = \lambda-\frac{1}{A}\expect{\Tint}\,.
\end{equation}
Likewise, we define corrected canonical single-particle energies by
\begin{equation}\label{eq:def_ecan_corr}
  \varepsilon^\text{corr}_\mu = h_{\mu\mu}-\frac{1}{A}\expect{\Tint}\,.
\end{equation}
In plain HF, the difference between the simple correction and the more involved approach amounts to $100-200\,\keV$ for levels near the Fermi surface. Likewise, a naive consideration of the exact separation energy $E_{N+1}-E_N$ would give rise to a correction to the canonical gap:
\begin{equation}
  \Delta^\text{corr}_\mu = \Delta_{\mu\bar\mu}-\frac{1}{A}\expect{\Tpair}\,,
\end{equation}
where the kinetic pairing energy $\expect{\Tpair}$ is typically $1-2\,\MeV$ at most in the tin isotopes, hence this correction would amount to $10-20\,\keV$ or less and has been omitted at present. The best strategy to avoid such conceptual uncertainties related to the perturbative definition of the gaps and single-particle energies is the self-consistent calculation of odd nuclei, which will be addressed in a subsequent publication. In addition, there are some general questions regarding the expectation values of $A$-dependent Hamiltonians in HFB states without sharp particle number that go beyond the scope of the present discussion and will be the studied elsewhere \cite{Hergert:2009}. These caveats should be kept in mind in the following sections.

\begin{figure}[t]
  \includegraphics[width=\columnwidth]{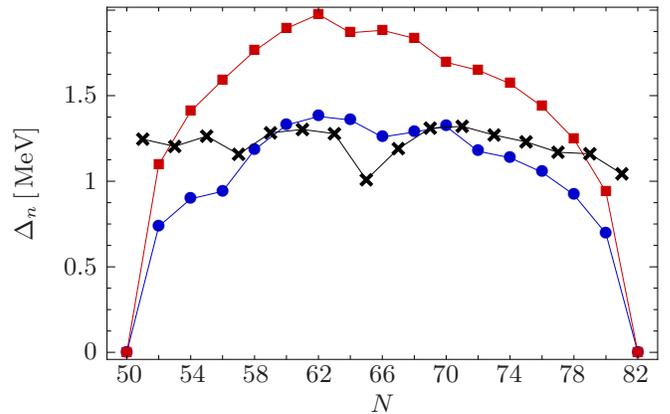}
  \caption{\label{fig:cmcomp}(Color online) Canonical neutron gaps of tin isotopes for the full intrinsic kinetic energy
  \symbolcircle[FGBlue]\, and the one-body approximation \symbolbox[FGRed]\, (see text). Experimental $\Delta^{(3)}(N)$ are indicated by \symbolxcross\, \cite{Audi:2002rp}. Calculations were done with the Gogny D1S interaction.}
\end{figure}

To conclude our discussion, we compare our treatment of the intrinsic kinetic energy to the widely used one-body approximation, which omits the two-body contribution in Eq. \eqref{eq:tint_obtb} altogether. Considering the tin isotopic chain, the ground-state energies and charge radii of the two approaches differ by 1\% at most (see Fig. \ref{fig:goge}), while the spectroscopic structure of the resulting ground states is somewhat different. Figure \ref{fig:ecmcomp} displays the canonical single-particle spectra; for the one-body approximation, further corrections to the single-particle energies are typically \emph{not} applied. The general trends of the single-particle energies are the same in both cases, but one notices that the calculations with the full intrinsic kinetic energy lead to a slightly reduced level density, which will impede pairing correlations and is expected to reduce the pairing energy or the gaps in comparison to the one-body approximation. Hence, we compare the canonical neutron gaps $\Delta_n$ of the two approaches in Fig. \ref{fig:cmcomp}. We find that the kinetic two-body term has a considerable effect, reducing the gap by as much as 30\%, i.e. about 500 $\keV$, for the mid-shell tin isotopes in calculations with the Gogny D1S interaction. This effect is due to the combination of the reduced level density and the repulsive kinetic energy contribution to the pairing field \eqref{eq:def_delta}. The actual size of the quenching will depend on the details of the fit of a phenomenological interaction, hence one cannot generalize the results for Gogny D1S easily to the Skyrme interactions, for instance. Comparing with the canonical single-particle spectrum, we also notice that the dips near the sub-shell closures in $\nuc{Sn}{106}, \nuc{Sn}{114}, \nuc{Sn}{120}$ are slightly enhanced in the calculation using the full intrinsic kinetic energy.

\section{\label{sec:pp_ucom}$\VUCOM$ as a pairing force}
\subsection{Theoretical Gap Systematics}
\begin{figure}[t]
  \includegraphics[width=\columnwidth]{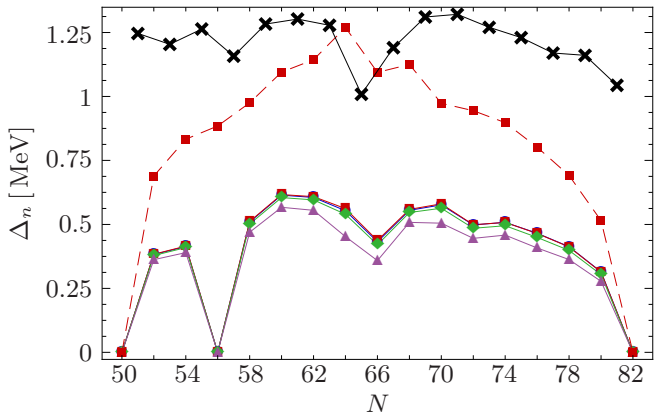}
  \caption{\label{fig:D1S_av18_srgA_alpha}(Color online) Canonical neutron gaps of tin isotopes for Gogny D1S+$\VUCOM$ with $\bar\alpha=0.03\,$\symbolcircle[FGBlue],$\;0.04\,$\symbolbox[FGRed],$\;0.06\,$\symboldiamond[FGGreen], and $0.1\,\fm^4$\symboltriangle[FGViolet]. Solid lines were obtained with the full $\Tint$, the dashed line with the one-body approximation. Experimental $\Delta^{(3)}(N)$ are indicated by \symbolxcross\, \cite{Audi:2002rp}.}
\end{figure}
To assess the pairing properties of $\VUCOM$, we perform hybrid calculations of the tin isotopic chain, using Gogny D1S in the particle-hole channel. Since the changes in ground-state energies and charge radii are minor compared to Fig. \ref{fig:goge}, we refrain from showing these results again, and focus directly on the canonical neutron pairing gaps. Figure \ref{fig:D1S_av18_srgA_alpha} shows the pairing gaps obtained using $\VUCOM$ for a range of parameters $\bar\alpha=0.03,\ldots,0.1\,\fm^4$. 

The canonical gaps obtained with the various $\VUCOM$ are about half the size of the ones obtained with Gogny D1S in the pairing channel (cf. Fig. \ref{fig:cmcomp}). In the mid-shell region, the experimental $\Delta^{(3)}(N)$ are underestimated by about 50\% as well. Varying the range of the UCOM transformation via $\bar\alpha$, we find that the gaps remain very stable for $\bar\alpha=0.03\,\fm^4$ to $0.06\,\fm^4$. Qualitatively, this implies that the attractive interaction matrix elements that are responsible for the pairing remain mostly unaffected in this range of parameters. This is indeed the case for the matrix elements in the relative ${}^1S_0$ partial wave \cite{Roth:2005pd,Hergert:2007wp,Roth:2008km}, which is expected to dominate the pairing at densities below saturation \cite{Dean:2002zx}. Moreover, the ${}^1S_0$ matrix elements hardly change at all beyond $\bar\alpha=0.05\,\fm^4$, and hence it is surprising that a notable reduction of the pairing gaps is found for $\bar\alpha=0.1\fm^4$, especially since $\VUCOM$ becomes more attractive overall at the same time (cf. Sect. \ref{sec:ucom}). 

To understand this observation, we have to consider two aspects of our calculations. First, we have to realize that the UCOM transformation causes a pre-diagonalization of the two-body Hamiltonian in momentum space, focusing the attractive and repulsive strength of the interaction near the diagonal region that is accessible in mean-field type calculations. Since the pairing is governed by the matrix elements near the Fermi surface, where the pairing tensor $\kappa$ is peaked, the pairing gaps are extremely sensitive to changes in the matrix elements in this very particular region of momentum space. The ground-state energy, however, is far less sensitive to such details. The second aspect is the formulation of the HFB method in a single-(quasi)particle basis, which implies that any two-nucleon state with total orbital angular momentum $L$ is a superposition of states with all allowed couplings of the center-of-mass and relative orbital angular momenta. This leads to admixtures of relative partial waves beyond ${}^1S_0$, which still exhibit a more significant $\bar\alpha$-dependence. Moreover, since the total spin of the nucleon pair is independent of the center-of-mass component of the two-body state, mostly repulsive spin-singlet relative partial waves are admixed to the ${}^1S_0$ wave, whereas the next significant contribution to pairing in nuclear matter is due to the spin-triplet ${}^3P_2$ partial waves (see, e.g., Ref. \cite{Dean:2002zx}).

\begin{figure}[t]
  \includegraphics[width=\columnwidth]{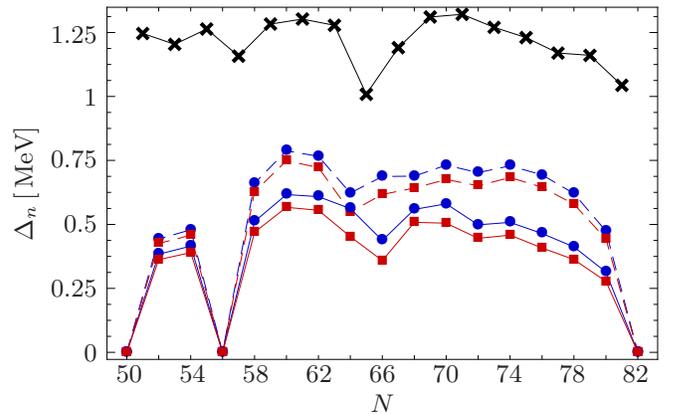}
  \caption{\label{fig:D1S_av18_srgA_pwave}(Color online) Canonical neutron gaps of tin isotopes for Gogny D1S+$\VUCOM$ with $\bar\alpha=0.04\,$\symbolcircle[FGBlue], and $0.1\,\fm^4$\symbolbox[FGRed]. Comparison of full interaction (solid) and ${}^1S_0$ partial wave (dashed) in the pairing channel. Experimental $\Delta^{(3)}(N)$ are indicated by \symbolxcross\,\cite{Audi:2002rp}.}
\end{figure}

For this reason, it is instructive to consider the effect of a partial-wave restriction for $\VUCOM$ in the pairing channel. The canonical neutron gaps obtained using merely the ${}^1S_0$ matrix elements of $\VUCOM$ with $\bar\alpha=0.04\,\fm^4$ and $\bar\alpha=0.1\,\fm^4$ are compared to those of the full calculations in Fig. \ref{fig:D1S_av18_srgA_pwave}. Whereas the results for the Gogny D1S are practically unaffected under such a restriction, we observe a significant effect of the higher partial waves in the case of $\VUCOM$, which cause a decrease of the gaps by as much as 20\%-30\% in the mid-shell tin isotopes. We also find that the difference between the restricted and full calculations is enhanced for the longer-ranged correlator $\bar\alpha=0.1\,\fm^4$: the inclusion of the higher partial waves reduces the gap by an additional 5\%-10\% compared to the calculation with $\bar\alpha=0.04\,\fm^4$. The different behavior of the two kinds of interactions can be explained by the comparably simple structure of Gogny D1S, which lacks $\qOV^2$, $\LOV^2$, and tensor terms that give rise to the more diverse partial wave structure of realistic NN interactions. Moreover, in the case of $\VUCOM$, one finds that the treatment of correlations induces a host of additional tensor operators (see, e.g., Ref. \cite{Roth:2005pd}).

\subsection{Comparison with SRG-evolved Interactions}
Recently, there has been an effort to use the RG-evolved low-momentum interaction $\Vlowk$ as a pairing interaction in conjunction with the Skyrme SLy4 force \cite{Duguet:2007be, Lesinski:2008cd}. Since this study is in the same spirit as the discussion in this section, we have carried out similar calculations using interactions obtained by evolving Argonne V18 via the SRG flow equation \eqref{eq:h_flow} to cutoffs
\begin{equation*}
  \lambda = \bar\alpha^{-1/4} = 1.8, 2.0, 2.4, 2.8 \fm^{-1}\,.
\end{equation*}
While $V_{\bar\alpha}$ and $\Vlowk$ are slightly different conceptually, a $\Vlowk$ with a soft cutoff function behaves very similar to an SRG-evolved interaction \cite{Bogner:2006pc,Bogner:2006vp}. Moreover, their properties with respect to binding energies in NCSM calculations or nuclear matter approaches are similar enough to expect the same for their pairing properties. Compared to $\VUCOM$, both $\Vlowk$ and $V_\text{SRG}$ are significantly softer, and require additional 3N forces to produce saturation in nuclear matter or finite nuclei (see Ref. \cite{Bogner:2007rx} and also Ref. \cite{Roth:2008km}). 

\begin{figure}[t]
  \includegraphics[width=\columnwidth]{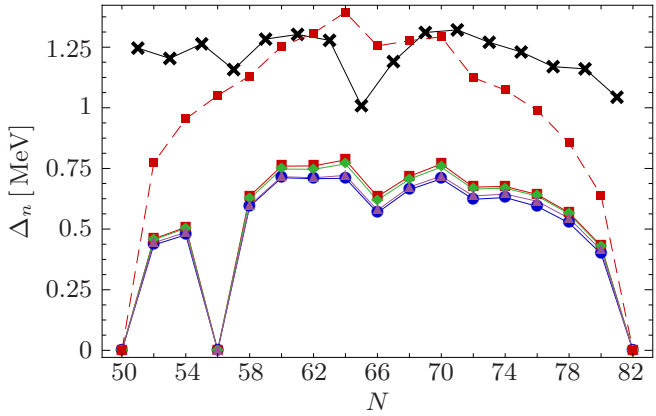}
  \caption{\label{fig:D1S_av18_srg_lambda}(Color online) Canonical neutron gaps of tin isotopes for Gogny D1S+$V_\text{SRG}$ with $\lambda=2.8\,$\symbolcircle[FGBlue],$\;2.4\,$\symbolbox[FGRed],$\;2.0\,$\symboldiamond[FGGreen], and $1.8\,\fm^{-1}$\symboltriangle[FGViolet]. Solid lines: full $\Tint$, dashed line: one-body approximation. Experimental $\Delta^{(3)}(N)$ are indicated by \symbolxcross\, \cite{Audi:2002rp}.}
\end{figure}

The resulting canonical neutron gaps of the tin chain are displayed in Fig. \ref{fig:D1S_av18_srg_lambda}. The cutoff variation leads to changes on the order of $10\%$ in the mid-shell region around $\nuc{Sn}{114}$. For $\lambda=2.0$ and $2.4\fm^{-1}$, the theoretical gaps are almost identical. For these values, the flow affects the interaction only at momenta that are already decoupled from the low-energy scales relevant for nuclear structure. The change obtained by lowering the cutoff from $2.8\,\fm^{-1}$ can then be understood as the shift of repulsion to higher momenta and many-nucleon terms, which is typical for the SRG evolution and renders the interaction more attractive in the partial waves that are relevant for pairing. As the $\VUCOM$ with $\bar\alpha=0.1\,\fm^4$ in the previous subsection, the corresponding SRG-evolved interaction with the lowest cutoff $1.8\,\fm^{-1}$ yields a decrease of the gaps.

The obtained results are considerably different from those of Ref. \cite{Lesinski:2008cd}, which presents canonical gaps $\Delta_n$ close to experimental gaps $\Delta^{(3)}(N)$ in a series of isotopic chains. Given our previous findings discussed in this paper, we argue that the possible reasons for this discrepancy are two-fold: as stated in \cite{Lesinski:2008cd}, Lesinski et al. use only the relative ${}^1S_0$ partial wave of $\Vlowk$ as an input in the particle-particle channel at present. In the previous subsection, we found that higher partial waves can reduce the canonical gaps by as much as 20\%-30\% (although in the specific case of $\VUCOM$). Moreover, in our discussion of the center-of-mass treatment in Sect. \ref{sec:kinetic_int} we observed a significant suppression of the gaps caused by the two-body term of the intrinsic kinetic energy (cf. Fig. \ref{fig:cmcomp}). For this reason, we have also included the gaps calculated with the one-body approximation to the kinetic energy for $\lambda=2.4\,\fm^{-1}$ in Fig. \ref{fig:D1S_av18_srg_lambda}. These gaps are indeed close to the experimental values, except near the major shell closures where one has to include pairing correlations beyond the HFB approximation. The suppression of the gaps due to the kinetic two-body term amounts to as much as 600-700$\keV$. A similar calculation for $\VUCOM$ with $\bar\alpha=0.04\,\fm^4$ is included in Fig. \ref{fig:D1S_av18_srgA_alpha} along with the results using the full $\Tint$, and it exhibits the same effect. 
We note, however, that our results were obtained with the specific choice \eqref{eq:tint_tb} for the intrinsic kinetic energy, whereas Lesinski et al. used the one- plus two-body form \eqref{eq:tint_obtb} \cite{Duguet:2009}. While Eqs. \eqref{eq:tint_tb} and \eqref{eq:tint_obtb} are equivalent at the operator level and it has been explicitly shown that they lead to the same energy expectation values in Hartree-Fock \cite{Khadkikar:1974}, it is not clear that this is still the case in HFB calculations, in particular due to use of an $A$-dependent Hamiltonian in a state without fixed particle number. We will analyze this issue in detail in a forthcoming publication \cite{Hergert:2009}. Until then, we cannot rule out that three-nucleon forces or beyond mean-field effects like the coupling to surface vibrations may have important effects on the pairing gaps.

At a first glance, our results are similar to those presented by Barranco et al. \cite{Barranco:2003kc, Pastore:2008zi} in studies on the impact of particle-vibration coupling on the pairing gap in $\nuc{Sn}{120}$, which combine the phenomenological Gogny D1S and SLy4 interactions in the particle-hole channel with the Argonne V14 interaction as a pairing force. We have to stress, however, that this agreement might be incidental because there are other important aspects that need to be considered \cite{Hebeler:2009dy,Lesinski:2009}. For instance, the use of the ``bare'', hard core AV14/18 in conjunction with phenomenological forces that are essentially of low-momentum character in the work of Barranco et al. is certainly a consistency issue. 

\section{\label{sec:ucom}Fully self-consistent HFB with $\VUCOM$}
\begin{figure}[t]
  \includegraphics[width=\columnwidth]{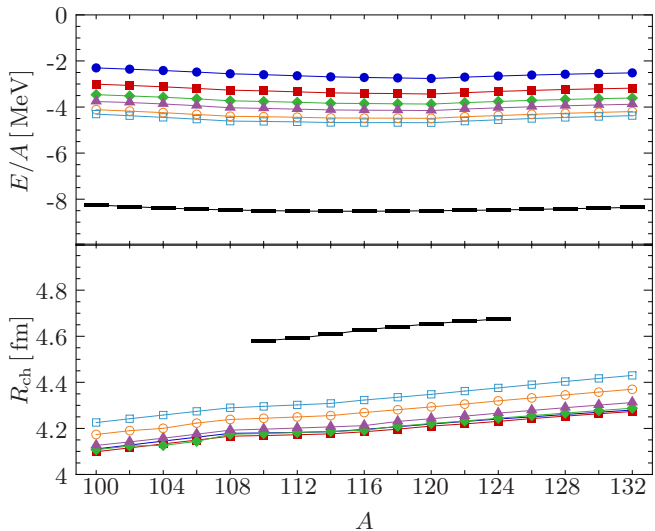}
  \caption{\label{fig:av18_srgAXXXX}(Color online) Ground-state energies (top) and charge radii (bottom) for the tin isotopes, calculated for $\VUCOM$ with 
    $\bar\alpha=0.03\,$\symbolcircle[FGBlue],\;$0.04\,$\symbolbox[FGRed],\;$0.05\,$\symboldiamond[FGGreen],\;$0.06\,$\symboltriangle[FGViolet],\;$0.08\,$\symbolcircleopen[FGOrange], and $0.1\,\fm^4\,$\symbolboxopen[FGLightBlue]. Experimental values \linemediumsolid\, taken from \cite{Audi:2002rp,DeVries:1987}.}
\end{figure}
Having gained some insight on how $\VUCOM$ behaves as a pairing force in the previous section, we now use a fully self-consistent HFB approach, using $\VUCOM$ in the particle-hole as well as the particle-particle channel. 

\subsection{Ground-State Energies and Radii}
We first consider the bulk properties of the tin isotopes obtained in a fully self-consistent calculation with $\VUCOM$ for $\bar\alpha=0.03,\ldots,\,1.0\,\fm^4$. Figure \ref{fig:av18_srgAXXXX} displays the resulting ground-state energies and charge radii. As expected from previous work, the nuclei are bound already at the mean-field level due to the explicit treatment of short-range correlations. The difference of $4$-$6\,\MeV$ per nucleon from experimental data is due to long-range correlations that are not described by the UCOM correlation operators and can be described by  beyond mean-field methods like many-body perturbation theory, as demonstrated successfully in \cite{Roth:2005ah,Guenther:2009}. The increase of the binding energy with $\bar\alpha$ implies that longer-ranged correlations are shifted from the many-body state into the correlation operators; it roughly corresponds to the cutoff dependence of results obtained with pure two-body $\Vlowk$ or SRG-evolved interactions.

In contrast to Ref. \cite{Roth:2005ah}, however, in which the UCOM transformation was constructed by a different strategy, the new SRG-generated correlation functions provide a significant improvement of the charge radii, which lie within roughly 10\% of experimental data and correctly reproduce the experimentally observed trends over the isotopic chain (see also \cite{Roth:2008km}). Since the radius is a long-ranged operator, it is rather insensitive to variations of $\bar\alpha$, as evident from Fig. \ref{fig:av18_srgAXXXX} (note the scale of the plot).

\subsection{Gaps}
\begin{figure}[t]
  \includegraphics[width=\columnwidth]{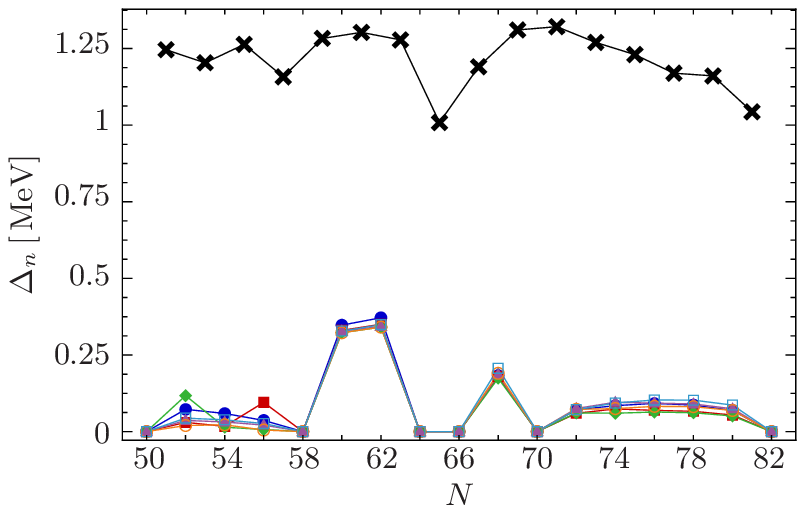}
  \caption{\label{fig:av18_srgA_alpha}(Color online) Canonical gaps in the tin isotopes for $\VUCOM$ with     $\bar\alpha=0.03\,$\symbolcircle[FGBlue],\;$0.04\,$\symbolbox[FGRed],\;$0.05\,$\symboldiamond[FGGreen],\;$0.06\,$\symboltriangle[FGViolet],\;$0.08\,$\symbolcircleopen[FGOrange], and $0.1\,\fm^4\,$\symbolboxopen[FGLightBlue], compared to experimental $\Delta^{(3)}(N)$ \,\symbolxcross\,\cite{Audi:2002rp}. }
\end{figure}

In Fig. \ref{fig:av18_srgA_alpha}, we show the canonical neutron gaps of the tin isotopes. As in the hybrid calculations in Sect. \ref{sec:pp_ucom}, the gaps are very stable under variations of $\bar\alpha$. It is noteworthy that this is even the case for the long-ranged $\VUCOM$ with $\bar\alpha=0.1\,\fm^4$. This improved stability could be a signal of the improved consistency, because unlike in the hybrid Gogny D1S+$\VUCOM$ calculations, the single-particle spectrum is directly affected by the variation of $\bar\alpha$ as well. Compared to the experimentally extracted $\Delta^{(3)}(N)$, the theoretical gaps are significantly lower, ranging from below $100\,\keV$ from the outer tin isotopes to roughly $400\,\keV$ in the mid-shell region. In addition, we find clear signals of each sub-shell closure in the tin nuclei.
 
\begin{figure}[t]
  \includegraphics[width=\columnwidth]{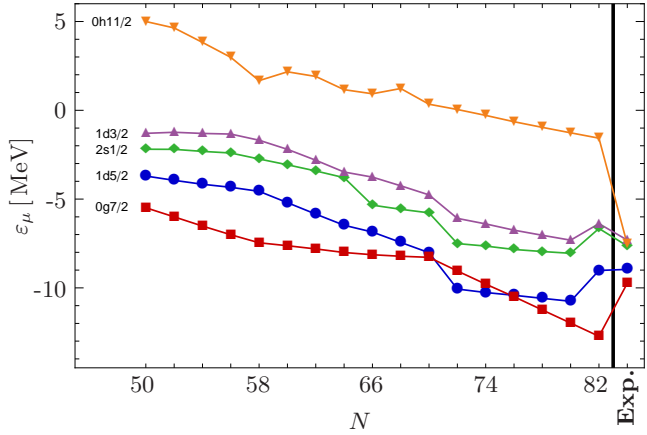}\\[-20pt]
  \caption{\label{fig:ecann}(Color online) Corrected canonical neutron single-particle energies in the tin isotopes, calculated for a $\VUCOM$ with $\bar\alpha=0.04\fm^4$. Experimental single-particle energies of $\nuc{Sn}{132}$ (``Exp.'') are included for reference \cite{Isakov:2002jv}.}
\end{figure}

 These findings can be understood if we consider the canonical single-particle spectra, which are shown for $\VUCOM$ with $\bar\alpha=0.04\,\fm^4$ in Fig. \ref{fig:ecann}. We note that the canonical neutron energies are spread over an interval on the order of $10\,\MeV$, which is about twice as large as for the purely phenomenological calculations with the Gogny D1S interaction in Fig. \ref{fig:ecmcomp}. The discrepancy between our spectra and the experimentally extracted single-particle levels of $\nuc{Sn}{132}$, which are included in Fig. \ref{fig:ecann} for reference, is even more severe. Such a low level density is a common feature of soft NN interactions, and consistent with previous studies using $\VUCOM$ \cite{Roth:2005ah, Paar:2006ua,Papakonstantinou:2006vc,Papakonstantinou:2008sf}. From BCS theory, it is well-known that the formation of Cooper pairs strongly depends on a sufficiently high level density in the region of the Fermi surface. For the BCS gap, this is reflected by the relation \cite{Bardeen:1957mv}
\begin{equation}
   \Delta \sim \exp{-\frac{1}{|g| n(0)}}
\end{equation}
where $g$ is the strength of the (attractive) pairing interaction and $n(0)$ the level density at the Fermi surface. Consequently, we see that the low level density generated by $\VUCOM$ presents a major obstacle to nuclear pairing. To obtain more realistic single-particle spectra, we will have to account for long-range correlations that are presently not described by either the correlation operators or the relatively simple many-body space, as well as three- or possibly higher many-nucleon forces.
 
 Comparing the theoretical and experimental single-particle levels for $\nuc{Sn}{132}$, we see that the $0g_{7/2}$ and $0h_{11/2}$ shells show a particularly large deviation, whereas the remaining levels and their splittings are reproduced rather well. Comparing with Fig. \ref{fig:av18_srgA_alpha}, we see that for these levels the canonical gaps are strongly suppressed as well, compared to the mid-shell region where the lowest canonical quasiparticle energies are associated with $s$ or $d$ orbitals. This strong dependence on the single-particle angular momenta suggests a significant influence of the tensor structure of $\VUCOM$. 

\begin{figure}[t]
  \includegraphics[width=\columnwidth]{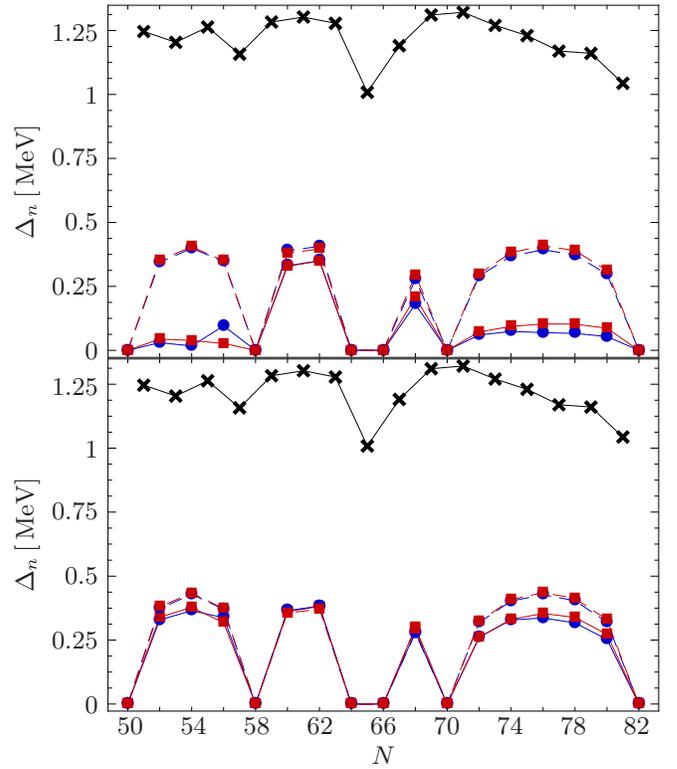}
  \caption{\label{fig:av18_srgA_pw}(Color online) Canonical (top) and average gaps (bottom) of tin isotopes for $\VUCOM$ with $\bar\alpha=0.04\,$\symbolcircle[FGBlue], and $0.1\,\fm^4$\symbolbox[FGRed]. Comparison of full interaction (solid) and ${}^1S_0$ partial wave (dashed). Experimental $\Delta^{(3)}(N)$ are indicated by \symbolxcross\, \cite{Audi:2002rp}.}
\end{figure}

To gain further insight, we first compare the canonical neutron gaps to the average gaps defined by Eq. \eqref{eq:def_avgDelta} in Fig. \ref{fig:av18_srgA_pw}. We find that the latter are practically constant over the tin isotopic chain, suggesting that the shells with low single-particle $j$ provide the essential contribution to this quantity, except at the sub-shell closures, where the pairing collapses and the solution is reduced to the HF case. Next, we revisit the restriction of the pairing interaction to the relative ${}^1S_0$ partial wave, which was discussed for the hybrid calculation in Sect. \ref{sec:pp_ucom}. This eliminates tensor effects in the pairing field, and as a result, we find a substantial increase in the canonical gaps in Fig. \ref{fig:av18_srgA_pw}. Considering that the same restriction of $\VUCOM$ in the Gogny D1S+$\VUCOM$ calculation presented in Fig. \ref{fig:D1S_av18_srgA_pwave} only caused a much smaller increase of the gap, we have to conclude that the interplay with the $ph$ interaction via self-consistency effects plays an important role as well. Furthermore, we see that without the tensor interaction, the canonical and average gaps are very similar, suggesting that the pairing is balanced more uniformly over all shells. 

Finally, we have tested the sensitivity of these results to changes in $\bar\alpha$. In Fig. \ref{fig:av18_srgA_pw}, we show only the calculation using the long-ranged $\VUCOM$ with $\bar\alpha=0.1\,\fm^4$ for comparison. As expected from the previous discussions, we find only minor differences due to the $\bar\alpha$-variation. The most notable changes occur in the high-$j$ subshells, and are in line with our previous statement that higher-lying partial waves require larger $\bar\alpha$ to become stable under $\bar\alpha$ variations (if at all). The negligible difference in both the canonical and average gaps when the pairing interaction is restricted to the relative ${}^1S_0$ waves serves as a further confirmation.  

The discussed observations underline the significance of the tensor structure of a realistic (albeit effective) NN interaction for the pairing correlations in finite nuclei. Although the HFB approach considers these effects merely on a mean-field level, and the possibility of a different behavior in more refined many-body methods exists, it should nevertheless be clear that the rather simplistic phenomenological forces obscure aspects of the two-nucleon physics that may prove to be very important for spectroscopic observables.
\section{Conclusions}
In this paper, we have presented a fully self-consistent HFB approach based on an intrinsic Hamiltonian. By using effective NN interactions derived from the realistic Argonne V18 interaction, we are able to maintain a stringent link to low-energy observables of the strong interaction, i.e., NN scattering phase shifts and deuteron properties. Since the same interactions can also be used in other Hartree-Fock based approaches and more refined ab-initio many-body methods like the No Core Shell Model \cite{Roth:2007sv} or the Coupled Cluster Method \cite{Roth:2008qd}, this opens important perspectives for cross-checking nuclear structure studies.

Focusing on the tin isotopic chain, we have calculated theoretical pairing gaps, and studied the effects of commonly used approximations in the center-of-mass treatment as well as the ansatz for phenomenological pairing interactions in Density Functional Theory \cite{Bender:2003jk}. We have discussed aspects of the tensorial structure of realistic NN interactions whose impact on the spectroscopic properties of finite nuclei is expected to be significant and that are described only inadequately by existing phenomenological functionals of the Skyrme or Gogny type \cite{Zalewski:2008is,Duguet:2008rr}. In particular, we have demonstrated that the inclusion of all partial waves of the NN interaction as well as the repulsive contribution of the intrinsic kinetic energy in the pairing field have a significant effect on the gaps. In the latter case, however, it remains to be seen whether this is affected by the particular choice of the intrinsic kinetic energy operator \cite{Duguet:2009}, and on the treatment of its $A$ dependence \cite{Hergert:2009}.

In a fully self-consistent approach using $\VUCOM$ in the particle-hole and the particle-particle channel, the low density of single-particle levels near the Fermi surface proves to be a strong impediment to nucleon Cooper pairing. This low level density, implying a low effective mass as well, is a general feature of soft, non-local interactions, and consistent with previous studies \cite{Roth:2005ah, Paar:2006ua}. Consequently, we conclude that beyond mean-field effects like the coupling to surface vibrations indeed play an important role in nuclear pairing. Such a coupling would lead to a dressing of the single-particle energies, and is expected to improve the level density near the Fermi surface, which would at least partially overcome the effects of the non-locality of $\VUCOM$ and similar interactions. 

In principle, there are two directions in which an extension of our framework can proceed: the inclusion of higher many-body forces, and the use of a more sophisticated many-body method. One of the ultimate aims of effective interaction methods is to obtain results that are independent of the control parameters of the transformation, in our case $\bar\alpha$. In this sense, the transformation merely yields a unitarily equivalent representation of low-energy QCD that is more suitable to the used many-body method. To achieve true consistency, however, one would need (i) to start from a consistent set of NN and higher nucleonic interactions, and (ii) to include them in the unitary transformation of the many-body states or operators. The former requirement can be met by using N3LO interactions derived in chiral effective field theory. While the required consistent set of NN, 3N, and 4N interactions have been worked out in principle (see, e.g., \cite{Bernard:2007sp, Epelbaum:2007us}), the complex structure of the full 3N terms and the computational demands for handling a 4N interaction have thus far prevented their use in many-body calculations. The second aspect complicates matters even further, because the inclusion of the 3N force in either UCOM or SRG transformations is a formidable challenge. At present, one therefore hopes that such a transformation may render the 3N interaction less important in actual calculations, so that a simpler model like the chiral N2LO interaction may reproduce the required effects after a readjustment of its parameters \cite{Nogga:2004ab}. In this spirit, we have studied the use of a regularized 3N contact force in conjunction with $\VUCOM$ as a first step \cite{Guenther:2009}. While initial HFB results including 3N forces together with first-generation UCOM interactions are available in \cite{Hergert:2008}, we are preparing a paper on such calculations with the new SRG-generated correlation functions discussed in Sect. \ref{sec:srg_ucom}.

As for the many-body methods, the use of more refined approximations improves the Hilbert space, enabling it to describe residual long-range correlations that are not explicitly treated by the UCOM correlation operators. In the context of HFB, a straightforward extension is the use of projection techniques to restore symmetries that are spontaneously broken in the calculated ground state. The simplest example is Particle Number Projection, which can be implemented rather easily, because the general structure of the HFB eigenvalue problem is preserved (see, e.g., \cite{Sheikh:1999yd}). Another approach that can describe additional correlations as well as collective behavior is the Quasiparticle Random Phase Approximation, using the HFB ground state as a starting point. Both of these methods were explored for the first-generation UCOM interactions in \cite{Hergert:2008}, and will be the subject of studies using the SRG-generated $\VUCOM$ interactions in subsequent papers.

\section*{Acknowledgments}
We thank T. Duguet for useful discussions. This work is supported by the Deutsche Forschungsgemeinschaft through contract SFB 634 and by the Helmholtz International Center for FAIR within the framework of the LOEWE program launched by the State of Hesse.

\appendix
\section{Expressions for the Gogny Interaction}
We briefly recall the parameterization of the Gogny interactions, which is given by \cite{Decharge:1980}
\begin{align}\label{eq:int-gogny}
    \vO_{12}&=\sum_{i=1}^2 \exp\left(-\frac{\rOV^2}{\mu_i^2}\right)
       \left(W_i+B_i\PO_\sigma-H_i\PO_\tau-M_i\PO_\sigma\PO_\tau\right) \notag\\
    &\hphantom{=}\quad+t_0\left(1+x_0\PO_\sigma\right)\rho^\alpha\left(\ROV\right)
        \delta^{3}\left(\rOV\right)\notag\\
    &\hphantom{=}\quad+iW_{LS}\left(\sigmaOV_1+\sigmaOV_2\right)\cdot
       \lnablaV\times\,\delta^{(3)}\left(\rOV\right)
       \rnablaV\,,
\end{align}
where $P_\sigma$ and $P_\tau$ are the usual spin and isospin exchange operators, e.g.,
\begin{equation}
  P_\sigma = \frac{1}{2}(1+\sigmasigmaO)\,,
\end{equation}
and the relative and center-of-mass positions are given by
\begin{align}
  \rOV&=\rOV_1-\rOV_2\,,\\[3pt]
  \ROV&=\half\left(\rOV_1+\rOV_2\right)\,,
\end{align}
and the gradient operator in the relative coordinates is
\begin{equation}
  \nablaV=\nablaV_1-\nablaV_2\,.
\end{equation}

\subsection{Two-Body Matrix Elements}
In Ref. \cite{Roth:2005ah}, we have provided expressions for the evaluation of two-body matrix elements in a relative $LS$-coupled spherical harmonic oscillator basis $\ket{n(LS)JMTM_T}$, and the subsequent Talmi transformation to obtain $jj$-coupled matrix elements for use in Hartree-Fock, Hartree-Fock-Bogoliubov and their extensions. With these expressions, the evaluation of two-body matrix elements of the finite range terms in Eq. \eqref{eq:int-gogny} is straightforward. 

While the spin-orbit interaction in \eqref{eq:int-gogny} is given in a form which facilitates the calculation of particle-hole and particle-particle fields for use in self-consistent field methods, the calculation of the corresponding relative two-body matrix element is somewhat more involved than for the usual relative $\spinorbitO$ interaction. Suppressing isospin indices as well as the angular-momentum projection $M$, and using rotational invariance, we obtain 
\begin{align}
&iW_{LS}
\matrixe{n(LS)J}{\left(\lnablaV\times\frac{\delta(\rO)}{4\pi\rO^2}\rnablaV\right)\cdot(\sigmaOV_1+\sigmaOV_2)}{n'(L'S')J} \notag\\
 &=-\frac{9}{\pi \aHO^2}W_{LS}(-1)^{J}\sixjsymb{1}{1}{1}{1}{1}{J} \notag\\
 &\qquad\qquad\times N_{n1}N_{n'1}\mathcal{L}^{3/2}_n(0)\mathcal{L}^{3/2}_{n'}(0)\delta_{L1}\delta_{LL'}
\delta_{S1}\,,
\end{align}
where $\mathcal{L}^{l+1/2}_n(x)$ are the Laguerre polynomials, $\aHO$ is the oscillator length of the relative basis, and 
\begin{equation}
  N_{nL}=\sqrt{\frac{2n!}{\aHO^3 \Gamma(n+L+\tfrac{3}{2})}}\,.
\end{equation}

The density-dependent matrix element is most conveniently evaluated in a $jj$-coupled basis, and one finds
\begin{align}\label{eq:ddi_me}
&\matrixe{n_1 l_1 j_1, n_2 l_2 j_2; J T}
     {\vO[\rho]}{n'_1 l'_1 j'_1, n'_2 l'_2 j'_2; J T}\notag\\[3pt]
&\qquad=\frac{1}{2}(1+(-1)^Tx_0)\frac{t_0\sqrt{\hat{j}_1\hat{j}_2\hat{j}'_1\hat{j}'_2}}{4\pi(2J+1)}
       \mathcal{I}_{n_1 l_1 n_2 l_2; n'_1 l'_1 n'_2 l'_2}\notag\\[3pt]
&\qquad\times
         \left\{\vphantom{\bigg|}\left(1-(-1)^{J+T+l_1+l_2}\right)(-1)^{j_2-j_2'+l_2+l_2'}\right.\notag\\
&\qquad\times \braket{j_1\half j_2-\half}{J0}\braket{j'_1\half j'_2-\half}{J0}\notag\\[3pt]
&\qquad\left.\vphantom{\bigg|}+\left(1+(-1)^T\vphantom{\Big|}\right)\braket{j_1\half j_2\half}{J1}\braket{j'_1\half j'_2\half}{J1}
\right\}\,,
\end{align}
with
\begin{align}
  \mathcal{I}_{n_1 l_1 n_2 l_2; n'_1 l'_1 n'_2 l'_2}&=\notag\\[3pt]
  \int dr\,r^2\rho^\alpha(r)R_{n_1l_1}(r)&R_{n_2l_2}(r)R_{n'_1l'_1}(r)R_{n'_2l'_2}(r)\,.
\end{align}
The angular-momentum and isospin projections have been suppressed since the matrix element does not depend on them.

\subsection{Fields}
For the sake of efficiency, we calculate the particle-hole and particle-particle fields of the density-dependent interaction as in density functional approaches rather than by contracting $\rho$ or $\kappa$ with the matrix element \eqref{eq:ddi_me} (see, e.g., Ref. \cite{Decharge:1980}). Using
\begin{align}
  \rho_\tau(\rV)&=\sum_{kk'}\rho^\tau_{kk'}\psi^*_{k\tau}(\rV)\psi_{k'\tau}(\rV)\,,\\
  \kappa_\tau(\rV)&=\sum_{k\bar{k}'}\kappa^\tau_{k\bar{k}'}\psi_{k\tau}(\rV)\psi_{\bar{k}'\tau}(\rV)\,,
\end{align}
where $\bar{k}$ denotes a time-reversed state and $\tau=p,n$, the matrix elements of the fields are given by
\begin{align}\label{eq:ddi_gme}
  \breve{\Gamma}^\tau_{kk'}&
                 =\int d^3r\,\psi^*_{k'\tau}(\rV)\breve{\Gamma}_\tau(\rV)\psi_{k\tau}(\rV)
\intertext{and}
  \breve{\Delta}^\tau_{kk'}&
              =\int d^3r\,\breve{\Delta}_\tau(\rV)\psi_{k\tau}(\rV)\psi_{k'\tau}(\rV)\label{eq:ddi_dme}\,.
\end{align}
Here,
\begin{align}
  \breve{\Gamma}_\tau(\rV)
           &=t_0\left[\left(1+\frac{x_0}{2}\right)\rho^{\alpha+1}(\rV)-
                                      \left(x_0+\frac{1}{2}\right)\rho^\alpha(\rV)\rho_\tau(\rV)\right]\notag\\
           &\hphantom{=}+
\frac{t_0}{4}\alpha\left(1-x_0\right)\rho^{\alpha-1}(\rV)
             \left(\rho^{2}(\rV)+\sum_{\tau'}|\kappa_{\tau'}(\rV)|^2\right)\notag\\
           &\hphantom{=}+
             t_0\alpha\left(x_0+\frac{1}{2}\right)\rho^{\alpha-1}(\rV)\rho_p(\rV)\rho_n(\rV)\,,
            \label{eq:ddi_g}
\end{align}
where the parts proportional to $\alpha$ constitute the rearrangement term due to the density dependence of the interaction, and
\begin{equation}\label{eq:ddi_d}
   \breve{\Delta}_\tau(\rV)=\frac{1}{2}t_0\left(1-x_0\right)\rho^\alpha(\rV)\kappa_\tau(\rV)\,.
\end{equation}

In the case of spherical symmetry, the densities are reduced to 
\begin{align}
  \rho_\tau(r)&=\sum_{ljnn'}\frac{2j+1}{4\pi}\rho^{lj\tau}_{nn'}R_{nl}(r)R_{n'l}(r)\,,\\
  \kappa_\tau(r)&=\sum_{\tau ljnn'}\frac{2j+1}{4\pi}(-1)^l\kappa^{lj\tau}_{nn'}R_{nl}(r)R_{n'l}(r)\,,
\end{align}
where the phase $(-1)^l$ appears due to using the properties of the spherical harmonics under time reversal, and $R_{nl}(r)$ are radial spherical harmonic oscillator wavefunctions. Likewise, the fields are
\begin{align}
  \breve{\Gamma}^{(lj\tau)}_{nn'}&=\int dr\,R_{nl}(r)\breve{\Gamma}_\tau(r)R_{n'l}(r)\,,\\
  \breve{\Delta}^{(lj\tau)}_{nn'}&=\int dr\,R_{nl}(r)\breve{\Delta}^{(lj)}_\tau(r)R_{n'l}(r)\,,
\end{align}
with
\begin{align}
  \breve{\Gamma}_\tau(r)
           &=t_0\left[\left(1+\frac{x_0}{2}\right)\rho^{\alpha+1}(r)-
                                      \left(x_0+\frac{1}{2}\right)\rho^\alpha(r)\rho_\tau(r)\right]\notag\\
           &\hphantom{=}+
\frac{t_0}{4}\alpha\left(1-x_0\right)\rho^{\alpha-1}(r)
             \left(\rho^2(r)+\sum_{\tau'}\kappa^2_{\tau'}(r)\right)\notag\\
           &\hphantom{=}+
             t_0\alpha\left(x_0+\frac{1}{2}\right)\rho^{\alpha-1}(r)\rho_p(r)\rho_n(r)\,,
\end{align}
where we have used that $\rho(r)$ and $\kappa(r)$ are real, and
\begin{equation}
   \breve{\Delta}^{(lj)}_\tau(r)=\frac{1}{2}t_0\left(1-x_0\right)\rho^\alpha(r)(-1)^l\kappa_\tau(r)\,.
\end{equation}


\end{document}